\documentclass{Interspeech}

\interspeechcameraready

\title{Probing the Robustness Properties of Neural Speech Codecs}

\author[affiliation={1}]{Wei-Cheng}{Tseng}
\author[affiliation={1}]{David}{Harwath}

\affiliation{Department of Computer Science}{University of Texas at Austin}{USA}
\email{raytseng@utexas.edu, harwath@utexas.edu}
\keywords{speech coding, neural speech codecs, robustness}

\usepackage{comment}
\usepackage{xcolor}
	
\usepackage{color, colortbl}
\definecolor{LightCyan}{rgb}{0.88,1,1}
\usepackage{pgfplots,pgfplotstable}
\pgfplotsset{compat=1.18} 
\usepackage{filecontents} 
\usepackage{graphicx}
\usepackage{subcaption}
\usepackage{multirow}
\usetikzlibrary{math}
\tikzmath
{
  function symlog(\x,\a){
    \yLarge = ((\x>\a) - (\x<-\a)) * (ln(max(abs(\x/\a),1)) + 1);
    \ySmall = (\x >= -\a) * (\x <= \a) * \x / \a ;
    return \yLarge + \ySmall ;
  };
  function symexp(\y,\a){
    \xLarge = ((\y>1) - (\y<-1)) * \a * exp(abs(\y) - 1) ;
    \xSmall = (\y>=-1) * (\y<=1) * \a * \y ;
    return \xLarge + \xSmall ;
  };
}

\usepackage{amssymb}
\usepackage{pifont}
\newcommand{\cmark}{\ding{51}}%
\newcommand{\xmark}{\ding{55}}%

\begin{document}

\maketitle

\begin{abstract}
Neural speech codecs have revolutionized speech coding, achieving higher compression while preserving audio fidelity. Beyond compression, they have emerged as tokenization strategies, enabling language modeling on speech and driving paradigm shifts across various speech processing tasks. Despite these advancements, their robustness in noisy environments remains underexplored, raising concerns about their generalization to real-world scenarios. In this work, we systematically evaluate neural speech codecs under various noise conditions, revealing non-trivial differences in their robustness. We further examine their linearity properties, uncovering non-linear distortions which partly explain observed variations in robustness. Lastly, we analyze their frequency response to identify factors affecting audio fidelity. Our findings provide critical insights into codec behavior and future codec design, as well as emphasizing the importance of noise robustness for their real-world integration\footnote{\scriptsize\url{ https://github.com/RayTzeng/Codec-Noise-Robustness}}.
\end{abstract}

\section{Introduction}
Compression of speech signals, formally known as speech coding, has been an active research area for decades. 
Early efforts focused on minimizing bitrates to reduce storage requirements while preserving intelligibility, playing a crucial role in digital communication systems~\cite{G722,MPEG4,valin2016high, o2023review}.
Recently, neural speech codecs~\cite{soundstream,Encodec,DAC, speechtokenizer,yang2023hifi,du2024funcodec,ai2024apcodec,parker2024scaling,liu2024semanticodec} have emerged as transformative approaches in this field. 
These models typically adopt autoencoder-based architectures, employing residual vector quantization~\cite{RVQ} (RVQ) to discretize high-dimensional latent representations into hierarchical sequences of codebook entries.
Trained on large-scale, diverse speech datasets, neural codecs achieve significantly higher compression ratios and superior audio fidelity compared to traditional signal processing methods~\cite{guo2025recent}.

Beyond compression, neural speech codecs have also proven valuable as tokenization strategies for speech signals, enabling streamlined speech-text joint modeling within unified frameworks~\cite{latif2023sparks,peng2024survey}.
This has sparked paradigm shifts in speech processing tasks, notably improving speech synthesis applications~\cite{VALL-E,viola,speechx,voicecraft,defossez2024moshi,zhan2024anygpt}.
Recent studies have further explored neural speech codecs as potential universal acoustic feature extractors, investigating whether their units could serve as an alternative to waveforms or mel spectrograms for general speech processing~\cite{viola,puvvada2024discrete,codecASR}.

Despite advancements in neural speech codecs, existing studies are often evaluated under their own specific experimental conditions, limiting the fairness of cross-comparisons and potentially overlooking practical limitations.
To address this, researchers have introduced standardized evaluation benchmarks facilitating systematic and comprehensive assessments across codecs from multiple perspectives.
For example:
Codec-SUPERB~\cite{codecSUPERB} assesses the information-preserving capabilities of neural speech codecs from downstream application perspectives.
DASB~\cite{mousavi2024dasb} examines the applicability of neural speech codec units across various speech processing tasks using simple probing heads.
Espnet-Codec~\cite{shi2024espnet} provides comprehensive evaluations across multiple metrics and multilingual datasets, offering holistic insights into codec performance.

However, we identify a critical aspect that has been largely overlooked: \textbf{the robustness of neural speech codecs in noisy environments}. 
Real-world speech signals often contain substantial background noise, but the robustness of neural codecs under such realistic and challenging conditions has rarely been systematically studied.
Unlike traditional codecs, which operate under predefined use cases with theoretically analyzable behavior, deep-learning-based neural codecs lack explicit interpretability. 
This opacity makes it difficult to anticipate failure modes, particularly in noisy and unseen acoustic scenarios.
As a result, these codecs may degrade unexpectedly in perceptual quality and propagate errors into downstream applications, adversely affecting the reliability of the resulting speech processing systems.

In this work, we systematically evaluate the noise robustness of neural speech codecs, providing deeper insights into their behavior. Our key contributions are as follows:
\begin{itemize}
    \item \textbf{Noise Robustness Evaluation}: We assess multiple neural speech codecs under various noise conditions, revealing non-trivial differences in robustness.
    We identify key factors influencing codec resilience, including training data diversity, bitrate selection, and quantization strategies.
    \item \textbf{Linearity Analysis}: 
    We investigate the practical linearity properties (additivity and homogeneity) of these codecs, uncovering non-linear distortions that may explain their behavior in realistic scenarios involving overlapped or complex audio signals.
    \item \textbf{Frequency Response Characterization}: We analyze the spectral distortions introduced by these codecs, illustrating how time-domain losses emphasize low frequencies while attenuating high ones, offering insights for improving codec design.
\end{itemize}
Through these analyses, our work aims to enhance the community’s understanding of neural codec robustness and encourage further research to improve their reliability and applicability in practical, noisy, real-world conditions.


\section{Preliminaries}
\subsection{Formulation of Compression and Reconstruction}
In this work, we consider neural speech codecs composed of three main components: an encoder $\mathbf{E}(\cdot)$, a RVQ module $\mathbf{RVQ}(\cdot)$, and a Decoder $\mathbf{D}(\cdot)$.
Let $X\in \mathbb{R}^T$ be an audio waveform sampled at rate $f_s$. The encoder first maps $X$ into a sequence of latent representations $Z=\mathbf{E}(X)\in\mathbb{R}^{N\times d}$, where $N$ is the number of frames and $d$ is the feature dimensionality.
Then, given $K$ codebooks (quantization stages), the RVQ module quantizes $Z$ into $K$ ordered sequences $Z_{quant}=\mathbf{RVQ}(Z)=[Z_1, Z_2, \cdots, Z_K$] by mapping the residual latent vector to one of the $C$ codebook entries at each stage.
Finally, $Z_{quant}$ is passed to the decoder, yielding the reconstructed signal $X'=\mathbf{D}(Z_{quant})$.
Note that the compressed speech signal can be represented by the mapped codebook entries during the residual quantization process.

\subsection{Definition of bitrate}
The bitrate of a codec quantifies the amount of information required to represent each second of speech signal.
Consider the speech signal $X$ described above. We define the frame rate (in frames per second) as $f_N=N\cdot f_s/T$. Then, given $k\in[K]$ quantization stages, the total bit per frame is $k\ \text{log}_2(C)$. Combining these, the bitrate $R$ (in bit per second) is defined as:
\begin{align}
    R=f_N\cdot k\ \text{log}_2(C)
\end{align}
Bitrate is a common measure of a codec model's compression efficiency. By selecting appropriate values of $k$ during inference, one can trade off between compression efficiency and reconstruction fidelity to accommodate different conditions. Note that this is a basic coding scheme and should be viewed as an upper bound on the bitrate. In practice, more efficient methods, such as entropy coding, can achieve a lower bitrate.

\section{Experimental Setups}
\begin{table}[]
\centering
\caption{Configurations of the neural speech codecs. S, A, M denote speech, audio (non-speech) and music, respectively.}
\renewcommand{\arraystretch}{0.6}
\setlength{\tabcolsep}{5pt}
\begin{tabular}{l|c|c|c|c|c}
\toprule
\multirow{2}{*}{\textbf{Model}}                     & \multicolumn{2}{c|}{\textbf{Configuration}} & \multicolumn{3}{c}{\textbf{Dataset}}\\
\cmidrule(l{1pt}r{1pt}){2-3}\cmidrule(l{1pt}r{1pt}){4-6}
& $f_s$~(Hz) & $R$~(kbps) & S & A & M\\
\midrule
Encodec~\cite{Encodec}                     & 24k     & 0.75-24 & \cmark & \cmark & \cmark  \\
\midrule
              DAC~\cite{DAC}              & 24k      & 0.75-24 & \cmark & \cmark & \cmark  \\
                            \midrule
SpeechTokenizer~\cite{speechtokenizer}             & 16k     & 4    & \cmark & \xmark & \xmark  \\
\midrule
                           HiFi-Codec~\cite{yang2023hifi} & 24k       & 3    & \cmark & \xmark & \xmark    \\
                            \midrule
FreqCodec~\cite{du2024funcodec}                   & 16k      & 0.5-16   & \cmark & \xmark & \xmark   \\
\bottomrule\bottomrule
\end{tabular}
\end{table}

Here, we describe the dataset used(sec.~\ref{sec:dataset}), the evaluated neural speech codecs(sec.~\ref{sec:codecs}), and the evaluation metrics(sec.~\ref{sec:metrics}).

\subsection{Datasets}
\label{sec:dataset}
In this work, we aim to systematically analyze the noise robustness and behavior of neural speech codecs. To ensure that the observed results are attributable to the codec models themselves rather than inherent noise in the dataset, we select the following clean speech datasets for evaluation:\\
\noindent\textbf{LibriSpeech}~\cite{panayotov2015librispeech} consists of approximately 1,000 hours of audiobook reading-style recordings.
The \texttt{test-clean} subset is used for evaluating perceptual quality (Mel Distance, PESQ), intelligibility (ASR-WER), and speaker identity (ASV-EER)\\
\noindent\textbf{RAVDESS}~\cite{livingstone2018ryerson} contains high-quality, expressive speech and song with rich emotional variations across eight categories. The \texttt{speech} subset is used for evaluating emotion preservation (SER-ACC).

\subsection{Neural speech codecs}
\label{sec:codecs}
We considered the following neural speech codecs in this work. Table 1 provides an overview of their configurations. \\
\noindent\textbf{Encodec}~\cite{Encodec} is a very widely used neural audio codec. It builds upon SoundStream~\cite{soundstream}, which utilizes an SEANet~\cite{tagliasacchi2020seanet} encoder-decoder architecture with RVQ. Encodec introduces a complementary LSTM~\cite{graves2012long} for enhanced sequence modeling of acoustic content and incorporates a multi-resolution mel loss to improve reconstruction quality. The model is trained on a mixture of speech, audio, and music datasets.\\
\noindent\textbf{DAC}~\cite{DAC} improved upon Encodec by mitigating codebook collapse with a modified training strategy. Additionally, it introduces a technique to address training-inference mismatch caused by quantizer dropout. The model is trained on a mixture of speech, audio and music datasets.\\
\noindent\textbf{SpeechTokenizer}~\cite{speechtokenizer} is a unified speech tokenizer predominantly designed for spoken language models. By distilling WavLM~\cite{wavlm}, the first codebook encodes semantic information of speech, while the residual quantization captures additional acoustic details. The model is trained on LibriSpeech~\cite{panayotov2015librispeech}.\\
\noindent\textbf{HiFi-Codec}~\cite{yang2023hifi} retaining high reconstruction quality at low bitrates. It introduces a novel Group-Residual Vector Quantization (GRVQ) method, which divides latent representations into multiple groups and quantizes each group independently using different RVQs. The model is trained on speech-only datasets.\\
\noindent\textbf{FreqCodec}~\cite{du2024funcodec}, which differs from other neural codecs that operate in the time domain, FreqCodec leverages frequency-domain features for compression. By quantizing spectral features instead of time-domain waveforms, it achieves comparable speech quality with reduced computational complexity. The model is trained on LibriTTS~\cite{libritts}.

\subsection{Evaluation Metrics}
\label{sec:metrics}
We employ objective metrics to evaluate reconstructed speech quality from multiple perspectives, focusing on perceptual quality and information-preservation crucial for downstream applications, acknowledging their growing adoption in speech processing tasks~\cite{VALL-E,viola,speechx,voicecraft,defossez2024moshi,zhan2024anygpt,puvvada2024discrete,codecASR}. \\
\noindent\textbf{Mel Distance}~\cite{MEL} measures the perceptual difference between the original and reconstructed speech in the frequency domain. Lower is better.\\
\noindent\textbf{PESQ} (Perceptual Evaluation of Speech Quality)~\cite{PESQ} predicts human perception of speech quality. Higher is better.\\
\noindent\textbf{WER} (Word Error Rate) on automatic speech recognition (ASR). A higher WER indicates greater distortion on content-related information. We use Whisper-Large~\cite{radford2023robust} in this study.\\
\noindent\textbf{EER} (Equal Error Rate) on automatic speaker verification (ASV). 
Higher EER reflects greater distortion on speaker characteristic. We employed ECAPA-TDNN~\cite{desplanques2020ecapa} in this study\footnote{https://github.com/NVIDIA/NeMo}.\\
\noindent\textbf{ACC} (Accuracy) from speech emotion recognition (SER). It indicates how well the reconstructed speech retains emotion-related information (e.g. timbre, prosodic contours). Higher is better. We use emotion2vec~\cite{ma2023emotion2vec} in this study.

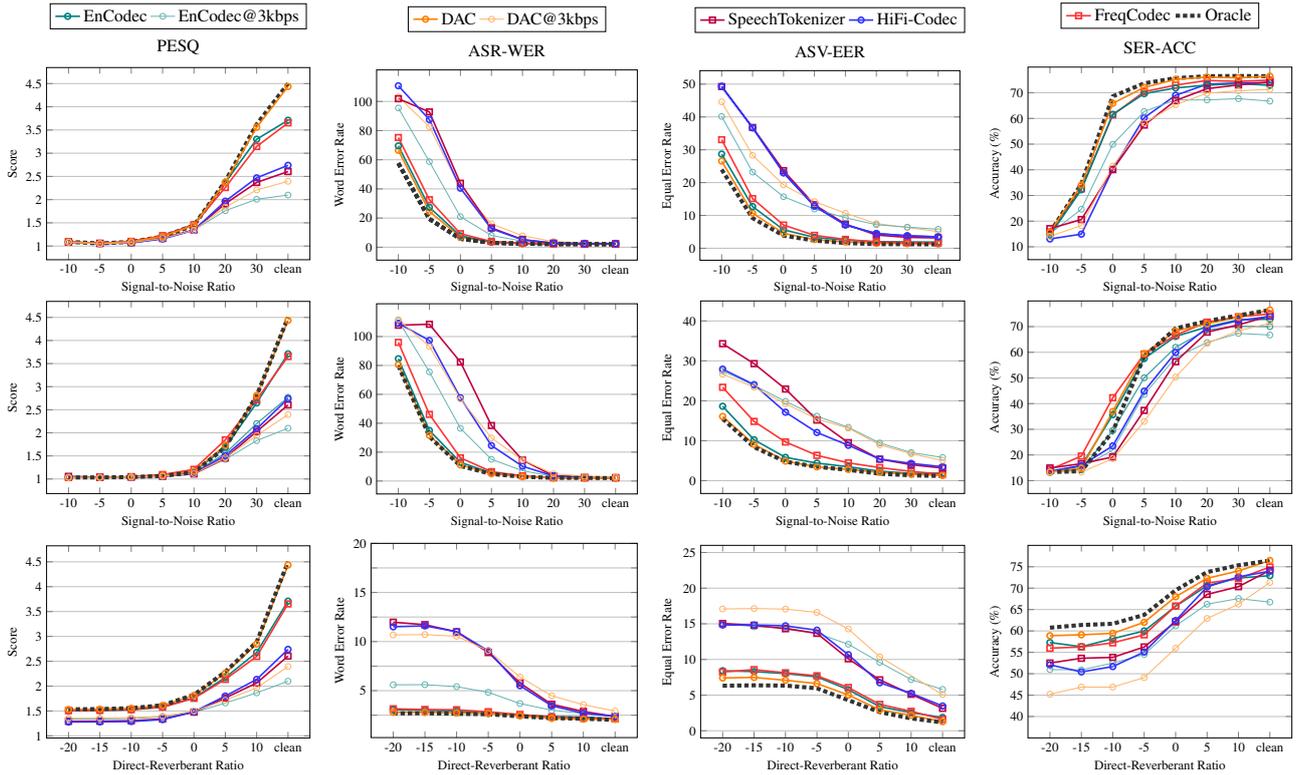
\begin{figure*}[!t]
    \centering
    \begin{subfigure}{0.24\textwidth}
        \centering
        \resizebox{\linewidth}{!}{\pgfplotstableread[col sep=comma,]{background-noise_pesq.csv}\datatable
\begin{tikzpicture}
\begin{axis}[
height=6.3cm,
width=8cm,
    title=PESQ,
    title style={font=\large},
    xtick=data,
    xticklabels from table={\datatable}{SNR},
    x tick label style={font=\normalsize},
    legend style={font=\large,at={(0.5,1.2)},anchor=south},
    legend columns=-1,
    ylabel={Score},
    xlabel={Signal-to-Noise Ratio},
    grid=major,
    ymajorgrids=true, 
    xmajorgrids=false,
    ytick = {1,1.5,2, 2.5, 3, 3.5,4,4.5},
    yticklabels={1,1.5,2, 2.5, 3, 3.5,4,4.5},
    ]

    \addplot [mark=o, teal!100, line width=0.4mm ] table [x expr=\coordindex, y={encodec_24k_24bps}]{\datatable};
    \addlegendentry{EnCodec}


    \addplot [mark=o, teal!60, line width=0.2mm ] table [x expr=\coordindex, y={encodec_24k_3bps}]{\datatable};
    \addlegendentry{EnCodec@3kbps}

    \addplot [black!80, line width=1.0mm, densely dashed ] table [x expr=\coordindex, y={oracle}]{\datatable};

    \addplot [mark=square, red!80, line width=0.4mm ] table [x expr=\coordindex, y={funcodec_en_libritts_16k_gr1nq32ds320}]{\datatable};

    \addplot [mark=square, purple!100, line width=0.4mm ] table [x expr=\coordindex, y={speech_tokenizer_16k}]{\datatable};
    

    \addplot [mark=o, blue!80, line width=0.4mm ] table [x expr=\coordindex, y={academicodec_hifi_24k_320d}]{\datatable};


    \addplot [mark=o, orange!100, line width=0.4mm ] table [x expr=\coordindex, y={dac_24khz}]{\datatable};

    \addplot [mark=o, orange!60, line width=0.2mm ] table [x expr=\coordindex, y={dac_24khz_4cb}]{\datatable};


    
    
    
    
    
    
\end{axis}
\end{tikzpicture}}
    \end{subfigure}
    \hfill
    \begin{subfigure}{0.24\textwidth}
        \centering
        \resizebox{\linewidth}{!}{\pgfplotstableread[col sep=comma,]{background-noise_wer.csv}\datatable
\begin{tikzpicture}
\begin{axis}[
height=6.3cm,
width=8cm,
    title=ASR-WER,
    title style={font=\large},
    xtick=data,
    xticklabels from table={\datatable}{SNR},
    x tick label style={font=\normalsize},
    legend style={font=\large,at={(0.5,1.2)},anchor=south},
    legend columns=-1,
    ylabel={Word Error Rate},
    xlabel={Signal-to-Noise Ratio},
    ytick = {0,20,40,60,80,100},
    yticklabels={0,20,40,60,80,100},
    grid=major,
    ymajorgrids=true, 
    xmajorgrids=false,
    ]

    \addplot [mark=o, orange!100, line width=0.4mm ] table [x expr=\coordindex, y={dac_24khz}]{\datatable};
    \addlegendentry{DAC}

    \addplot [mark=o, orange!60, line width=0.2mm ] table [x expr=\coordindex, y={dac_24khz_4cb}]{\datatable};
    \addlegendentry{DAC@3kbps}

    \addplot [mark=o, teal!60, line width=0.2mm ] table [x expr=\coordindex, y={encodec_24k_3bps}]{\datatable};

    \addplot [black!80, line width=1.0mm, densely dashed ] table [x expr=\coordindex, y={oracle}]{\datatable};

    \addplot [mark=o, teal!100, line width=0.4mm ] table [x expr=\coordindex, y={encodec_24k_24bps}]{\datatable};




    \addplot [black!80, line width=1.0mm, densely dashed ] table [x expr=\coordindex, y={oracle}]{\datatable};

    \addplot [mark=square, red!80, line width=0.4mm ] table [x expr=\coordindex, y={funcodec_en_libritts_16k_gr1nq32ds320}]{\datatable};

    \addplot [mark=square, purple!100, line width=0.4mm ] table [x expr=\coordindex, y={speech_tokenizer_16k}]{\datatable};
    

    \addplot [mark=o, blue!80, line width=0.4mm ] table [x expr=\coordindex, y={academicodec_hifi_24k_320d}]{\datatable};



    \addplot [black!80, line width=1.0mm, densely dashed ] table [x expr=\coordindex, y={oracle}]{\datatable};


    
    
    
    
    
    
\end{axis}
\end{tikzpicture}}
    \end{subfigure}
    \hfill
    \begin{subfigure}{0.24\textwidth}
        \centering
        \resizebox{\linewidth}{!}{\pgfplotstableread[col sep=comma,]{background-noise_eer.csv}\datatable
\begin{tikzpicture}
\begin{axis}[
height=6.3cm,
width=8cm,
title=ASV-EER,
title style={font=\large},
    xtick=data,
    xticklabels from table={\datatable}{SNR},
    x tick label style={font=\normalsize},
    legend style={font=\large,at={(0.5,1.2)},anchor=south},
    legend columns=-1,
    ylabel={Equal Error Rate},
    xlabel={Signal-to-Noise Ratio},
    grid=major,
    ymajorgrids=true, 
    xmajorgrids=false,
    ytick = {0,10,20,30,40,50},
    yticklabels={0,10,20,30,40,50},
    ]

    \addplot [mark=square, purple!100, line width=0.4mm ] table [x expr=\coordindex, y={speech_tokenizer_16k}]{\datatable};
    \addlegendentry{SpeechTokenizer}

    \addplot [mark=o, blue!80, line width=0.4mm ] table [x expr=\coordindex, y={academicodec_hifi_24k_320d}]{\datatable};
    \addlegendentry{HiFi-Codec}

    \addplot [mark=o, orange!100, line width=0.4mm ] table [x expr=\coordindex, y={dac_24khz}]{\datatable};

    \addplot [mark=o, orange!60, line width=0.2mm ] table [x expr=\coordindex, y={dac_24khz_4cb}]{\datatable};

    \addplot [mark=o, teal!100, line width=0.4mm ] table [x expr=\coordindex, y={encodec_24k_24bps}]{\datatable};


    \addplot [mark=o, teal!60, line width=0.2mm ] table [x expr=\coordindex, y={encodec_24k_3bps}]{\datatable};

    \addplot [mark=square, red!80, line width=0.4mm ] table [x expr=\coordindex, y={funcodec_en_libritts_16k_gr1nq32ds320}]{\datatable};


    \addplot [mark=o, blue!80, line width=0.4mm ] table [x expr=\coordindex, y={academicodec_hifi_24k_320d}]{\datatable};




    \addplot [black!80, line width=1.0mm, densely dashed ] table [x expr=\coordindex, y={oracle}]{\datatable};


    
    
    
    
    
    
\end{axis}
\end{tikzpicture}}
    \end{subfigure}
    \hfill
    \begin{subfigure}{0.24\textwidth}
        \centering
        \resizebox{\linewidth}{!}{\pgfplotstableread[col sep=comma,]{background-noise_ser.csv}\datatable
\begin{tikzpicture}
\begin{axis}[
height=6.3cm,
width=8cm,
    title=SER-ACC,
    title style={font=\large},
    xtick=data,
    xticklabels from table={\datatable}{SNR},
    x tick label style={font=\normalsize},
    legend style={font=\large,at={(0.5,1.2)},anchor=south},
    legend columns=-1,
    ylabel={Accuracy (\%)},
    xlabel={Signal-to-Noise Ratio},
    ytick = {10,20,30,40,50,60,70},
    yticklabels={10,20,30,40,50,60,70},
    grid=major,
    ymajorgrids=true, 
    xmajorgrids=false,
     ymin=5,ymax=80]


    \addplot [mark=square, red!80, line width=0.4mm ] table [x expr=\coordindex, y={funcodec_en_libritts_16k_gr1nq32ds320}]{\datatable};
    \addlegendentry{FreqCodec}

    \addplot [black!80, line width=1.0mm, densely dashed ] table [x expr=\coordindex, y={oracle}]{\datatable};
    \addlegendentry{Oracle}

    \addplot [mark=o, blue!80, line width=0.4mm ] table [x expr=\coordindex, y={academicodec_hifi_24k_320d}]{\datatable};

    \addplot [mark=o, teal!100, line width=0.4mm ] table [x expr=\coordindex, y={encodec_24k_24bps}]{\datatable};


    \addplot [mark=o, teal!60, line width=0.2mm ] table [x expr=\coordindex, y={encodec_24k_3bps}]{\datatable};

    \addplot [mark=square, purple!100, line width=0.4mm ] table [x expr=\coordindex, y={speech_tokenizer_16k}]{\datatable};
    


    \addplot [mark=o, orange!100, line width=0.4mm ] table [x expr=\coordindex, y={dac_24khz}]{\datatable};

    \addplot [mark=o, orange!60, line width=0.2mm ] table [x expr=\coordindex, y={dac_24khz_4cb}]{\datatable};



    
    
    
    
    
    
\end{axis}
\end{tikzpicture}}
    \end{subfigure}
    \hfill
    \begin{subfigure}{0.24\textwidth}
        \centering
        \resizebox{\linewidth}{!}{\pgfplotstableread[col sep=comma,]{white-noise_pesq.csv}\datatable
\begin{tikzpicture}
\begin{axis}[
height=6.3cm,
width=8cm,
    title style={font=\large},
    xtick=data,
    xticklabels from table={\datatable}{SNR},
    x tick label style={font=\normalsize},
    legend style={font=\large,at={(0.02,0.98)},anchor=north west},
    ylabel={Score},
    xlabel={Signal-to-Noise Ratio},
    grid=major,
    ymajorgrids=true, 
    xmajorgrids=false,
    ytick = {1,1.5,2, 2.5, 3, 3.5,4,4.5},
    yticklabels={1,1.5,2, 2.5, 3, 3.5,4,4.5},
    ]

    \addplot [mark=o, teal!100, line width=0.4mm ] table [x expr=\coordindex, y={encodec_24k_24bps}]{\datatable};

    \addplot [mark=o, teal!75, line width=0.3mm ] table [x expr=\coordindex, y={encodec_24k_6bps}]{\datatable};

    \addplot [mark=o, teal!60, line width=0.2mm ] table [x expr=\coordindex, y={encodec_24k_3bps}]{\datatable};

    \addplot [mark=square, red!80, line width=0.4mm ] table [x expr=\coordindex, y={funcodec_en_libritts_16k_gr1nq32ds320}]{\datatable};

    \addplot [mark=square, purple!100, line width=0.4mm ] table [x expr=\coordindex, y={speech_tokenizer_16k}]{\datatable};
    

    \addplot [mark=o, blue!80, line width=0.4mm ] table [x expr=\coordindex, y={academicodec_hifi_24k_320d}]{\datatable};


    \addplot [mark=o, orange!100, line width=0.4mm ] table [x expr=\coordindex, y={dac_24khz}]{\datatable};

    \addplot [mark=o, orange!60, line width=0.2mm ] table [x expr=\coordindex, y={dac_24khz_4cb}]{\datatable};

    \addplot [black!80, line width=1.0mm, densely dashed ] table [x expr=\coordindex, y={oracle}]{\datatable};


    
    
    
    
    
    
\end{axis}
\end{tikzpicture}}
    \end{subfigure}
    \hfill
    \begin{subfigure}{0.24\textwidth}
        \centering
        \resizebox{\linewidth}{!}{\pgfplotstableread[col sep=comma,]{white-noise_wer.csv}\datatable
\begin{tikzpicture}
\begin{axis}[
height=6.3cm,
width=8cm,
    title style={font=\large},
    xtick=data,
    xticklabels from table={\datatable}{SNR},
    x tick label style={font=\normalsize},
    legend style={font=\large,at={(0.98,0.98)},anchor=north east},
    ylabel={Word Error Rate},
    xlabel={Signal-to-Noise Ratio},
    ytick = {0,20,40,60,80,100},
    yticklabels={0,20,40,60,80,100},
    grid=both,
    ymajorgrids=true, 
    xmajorgrids=false,
    ]

    \addplot [mark=o, teal!100, line width=0.4mm ] table [x expr=\coordindex, y={encodec_24k_24bps}]{\datatable};


    \addplot [mark=o, teal!60, line width=0.2mm ] table [x expr=\coordindex, y={encodec_24k_3bps}]{\datatable};

    \addplot [mark=square, red!80, line width=0.4mm ] table [x expr=\coordindex, y={funcodec_en_libritts_16k_gr1nq32ds320}]{\datatable};

    \addplot [mark=square, purple!100, line width=0.4mm ] table [x expr=\coordindex, y={speech_tokenizer_16k}]{\datatable};
    

    \addplot [mark=o, blue!80, line width=0.4mm ] table [x expr=\coordindex, y={academicodec_hifi_24k_320d}]{\datatable};


    \addplot [mark=o, orange!100, line width=0.4mm ] table [x expr=\coordindex, y={dac_24khz}]{\datatable};

    \addplot [mark=o, orange!60, line width=0.2mm ] table [x expr=\coordindex, y={dac_24khz_4cb}]{\datatable};

    \addplot [black!80, line width=1.0mm, densely dashed ] table [x expr=\coordindex, y={oracle}]{\datatable};


    
    
    
    
    
    
\end{axis}
\end{tikzpicture}}
    \end{subfigure}
    \hfill
    \begin{subfigure}{0.24\textwidth}
        \centering
        \resizebox{\linewidth}{!}{\pgfplotstableread[col sep=comma,]{white-noise_eer.csv}\datatable
\begin{tikzpicture}
\begin{axis}[
height=6.3cm,
width=8cm,
    xtick=data,
    xticklabels from table={\datatable}{SNR},
    x tick label style={font=\normalsize},
    legend style={at={(0.02,0.98)},anchor=north west},
    ylabel={Equal Error Rate},
    xlabel={Signal-to-Noise Ratio},
    grid=major,
    ymajorgrids=true, 
    xmajorgrids=false,
    ymax=45,
    ytick = {0,10,20,30,40},
    yticklabels={0,10,20,30,40},
    ]

    \addplot [mark=o, teal!100, line width=0.4mm ] table [x expr=\coordindex, y={encodec_24k_24bps}]{\datatable};


    \addplot [mark=o, teal!60, line width=0.2mm ] table [x expr=\coordindex, y={encodec_24k_3bps}]{\datatable};

    \addplot [mark=square, red!80, line width=0.4mm ] table [x expr=\coordindex, y={funcodec_en_libritts_16k_gr1nq32ds320}]{\datatable};

    \addplot [mark=square, purple!100, line width=0.4mm ] table [x expr=\coordindex, y={speech_tokenizer_16k}]{\datatable};
    

    \addplot [mark=o, blue!80, line width=0.4mm ] table [x expr=\coordindex, y={academicodec_hifi_24k_320d}]{\datatable};


    \addplot [mark=o, orange!100, line width=0.4mm ] table [x expr=\coordindex, y={dac_24khz}]{\datatable};

    \addplot [mark=o, orange!60, line width=0.2mm ] table [x expr=\coordindex, y={dac_24khz_4cb}]{\datatable};

    \addplot [black!80, line width=1.0mm, densely dashed ] table [x expr=\coordindex, y={oracle}]{\datatable};


    
    
    
    
    
    
\end{axis}
\end{tikzpicture}}
    \end{subfigure}
    \hfill
    \begin{subfigure}{0.24\textwidth}
        \centering
        \resizebox{\linewidth}{!}{\pgfplotstableread[col sep=comma,]{white-noise_ser.csv}\datatable
\begin{tikzpicture}
\begin{axis}[
height=6.3cm,
width=8cm,
    title style={font=\large},
    xtick=data,
    xticklabels from table={\datatable}{SNR},
    x tick label style={font=\normalsize},
    legend style={font=\large,at={(0.98,0.98)},anchor=north east},
    ylabel={Accuracy (\%)},
    xlabel={Signal-to-Noise Ratio},
   ytick = {10,20,30,40,50,60,70},
    yticklabels={10,20,30,40,50,60,70},
    grid=major,
    ymajorgrids=true, 
    xmajorgrids=false,
    ymin=5,ymax=80]

    \addplot [mark=o, teal!100, line width=0.4mm ] table [x expr=\coordindex, y={encodec_24k_24bps}]{\datatable};

    \addplot [mark=o, teal!75, line width=0.3mm ] table [x expr=\coordindex, y={encodec_24k_6bps}]{\datatable};

    \addplot [mark=o, teal!60, line width=0.2mm ] table [x expr=\coordindex, y={encodec_24k_3bps}]{\datatable};

    \addplot [mark=square, red!80, line width=0.4mm ] table [x expr=\coordindex, y={funcodec_en_libritts_16k_gr1nq32ds320}]{\datatable};

    \addplot [mark=square, purple!100, line width=0.4mm ] table [x expr=\coordindex, y={speech_tokenizer_16k}]{\datatable};
    

    \addplot [mark=o, blue!80, line width=0.4mm ] table [x expr=\coordindex, y={academicodec_hifi_24k_320d}]{\datatable};


    \addplot [mark=o, orange!100, line width=0.4mm ] table [x expr=\coordindex, y={dac_24khz}]{\datatable};

    \addplot [mark=o, orange!60, line width=0.2mm ] table [x expr=\coordindex, y={dac_24khz_4cb}]{\datatable};

    \addplot [black!80, line width=1.0mm, densely dashed ] table [x expr=\coordindex, y={oracle}]{\datatable};


    
    
    
    
    
    
\end{axis}
\end{tikzpicture}}
    \end{subfigure}
    \hfill
    \begin{subfigure}{0.24\textwidth}
        \centering
        \resizebox{\linewidth}{!}{\pgfplotstableread[col sep=comma,]{reverb_pesq.csv}\datatable
\begin{tikzpicture}
\begin{axis}[
height=6.3cm,
width=8cm,
    title style={font=\large},
    xtick=data,
    xticklabels from table={\datatable}{SNR},
    x tick label style={font=\normalsize},
    legend style={font=\large,at={(0.02,0.98)},anchor=north west},
    ylabel={Score},
    xlabel={Direct-Reverberant Ratio},
    grid=major,
    ymajorgrids=true, 
    xmajorgrids=false,
    ytick = {1,1.5,2, 2.5, 3, 3.5,4,4.5},
    yticklabels={1,1.5,2, 2.5, 3, 3.5,4,4.5},
    ]

    \addplot [mark=o, teal!100, line width=0.4mm ] table [x expr=\coordindex, y={encodec_24k_24bps}]{\datatable};


    \addplot [mark=o, teal!60, line width=0.2mm ] table [x expr=\coordindex, y={encodec_24k_3bps}]{\datatable};

    \addplot [mark=square, red!80, line width=0.4mm ] table [x expr=\coordindex, y={funcodec_en_libritts_16k_gr1nq32ds320}]{\datatable};

    \addplot [mark=square, purple!100, line width=0.4mm ] table [x expr=\coordindex, y={speech_tokenizer_16k}]{\datatable};
    

    \addplot [mark=o, blue!80, line width=0.4mm ] table [x expr=\coordindex, y={academicodec_hifi_24k_320d}]{\datatable};


    \addplot [mark=o, orange!100, line width=0.4mm ] table [x expr=\coordindex, y={dac_24khz}]{\datatable};

    \addplot [mark=o, orange!60, line width=0.2mm ] table [x expr=\coordindex, y={dac_24khz_4cb}]{\datatable};

    \addplot [black!80, line width=1.0mm, densely dashed ] table [x expr=\coordindex, y={oracle}]{\datatable};


    
    
    
    
    
    
\end{axis}
\end{tikzpicture}}
    \end{subfigure}
    \hfill
    \begin{subfigure}{0.24\textwidth}
        \centering
        \resizebox{\linewidth}{!}{\pgfplotstableread[col sep=comma,]{reverb_wer.csv}\datatable
\begin{tikzpicture}
\begin{axis}[
height=6.3cm,
width=8cm,
    title style={font=\large},
    xtick=data,
    xticklabels from table={\datatable}{SNR},
    x tick label style={font=\normalsize},
    legend style={font=\large,at={(0.98,0.98)},anchor=north east},
    ylabel={Word Error Rate},
    xlabel={Direct-Reverberant Ratio},
    ytick = {2.5,5,7.5,10,12.5,15,17.5,20},
    yticklabels={,5,,10,,15,,20},
    grid=both,
    ymajorgrids=true, 
    xmajorgrids=false,
    ymax=20
    ]

    \addplot [mark=o, teal!100, line width=0.4mm ] table [x expr=\coordindex, y={encodec_24k_24bps}]{\datatable};


    \addplot [mark=o, teal!60, line width=0.2mm ] table [x expr=\coordindex, y={encodec_24k_3bps}]{\datatable};

    \addplot [mark=square, red!80, line width=0.4mm ] table [x expr=\coordindex, y={funcodec_en_libritts_16k_gr1nq32ds320}]{\datatable};

    \addplot [mark=square, purple!100, line width=0.4mm ] table [x expr=\coordindex, y={speech_tokenizer_16k}]{\datatable};
    

    \addplot [mark=o, blue!80, line width=0.4mm ] table [x expr=\coordindex, y={academicodec_hifi_24k_320d}]{\datatable};


    \addplot [mark=o, orange!100, line width=0.4mm ] table [x expr=\coordindex, y={dac_24khz}]{\datatable};

    \addplot [mark=o, orange!60, line width=0.2mm ] table [x expr=\coordindex, y={dac_24khz_4cb}]{\datatable};

    \addplot [black!80, line width=1.0mm, densely dashed ] table [x expr=\coordindex, y={oracle}]{\datatable};


    
    
    
    
    
    
\end{axis}
\end{tikzpicture}}
    \end{subfigure}
    \hfill
    \begin{subfigure}{0.24\textwidth}
        \centering
        \resizebox{\linewidth}{!}{\pgfplotstableread[col sep=comma,]{reverb_eer.csv}\datatable
\begin{tikzpicture}
\begin{axis}[
height=6.3cm,
width=8cm,
    xtick=data,
    xticklabels from table={\datatable}{SNR},
    x tick label style={font=\normalsize},
    legend style={at={(0.02,0.98)},anchor=north west},
    ylabel={Equal Error Rate},
    xlabel={Direct-Reverberant Ratio},
    grid=major,
    ymajorgrids=true, 
    xmajorgrids=false,
    ytick = {0,5,10,15,20,25},
    yticklabels={0,5,10,15,20,25},
    ymin=-1,ymax=26
    ]

    \addplot [mark=o, teal!100, line width=0.4mm ] table [x expr=\coordindex, y={encodec_24k_24bps}]{\datatable};


    \addplot [mark=o, teal!60, line width=0.2mm ] table [x expr=\coordindex, y={encodec_24k_3bps}]{\datatable};

    \addplot [mark=square, red!80, line width=0.4mm ] table [x expr=\coordindex, y={funcodec_en_libritts_16k_gr1nq32ds320}]{\datatable};

    \addplot [mark=square, purple!100, line width=0.4mm ] table [x expr=\coordindex, y={speech_tokenizer_16k}]{\datatable};
    

    \addplot [mark=o, blue!80, line width=0.4mm ] table [x expr=\coordindex, y={academicodec_hifi_24k_320d}]{\datatable};


    \addplot [mark=o, orange!100, line width=0.4mm ] table [x expr=\coordindex, y={dac_24khz}]{\datatable};

    \addplot [mark=o, orange!60, line width=0.2mm ] table [x expr=\coordindex, y={dac_24khz_4cb}]{\datatable};

    \addplot [black!80, line width=1.0mm, densely dashed ] table [x expr=\coordindex, y={oracle}]{\datatable};


    
    
    
    
    
    
\end{axis}
\end{tikzpicture}}
    \end{subfigure}
    \hfill
    \begin{subfigure}{0.24\textwidth}
        \centering
        \resizebox{\linewidth}{!}{\pgfplotstableread[col sep=comma,]{reverb_ser.csv}\datatable
\begin{tikzpicture}
\begin{axis}[
height=6.3cm,
width=8cm,
    title style={font=\large},
    xtick=data,
    xticklabels from table={\datatable}{SNR},
    x tick label style={font=\normalsize},
    legend style={font=\large,at={(0.98,0.98)},anchor=north east},
    ylabel={Accuracy (\%)},
    xlabel={Direct-Reverberant Ratio},
    ytick = {40,45,50,55,60,65,70,75},
    yticklabels={40,45,50,55,60,65,70,75},
    grid=major,
    ymajorgrids=true, 
    xmajorgrids=false,
    ymin=35,ymax=80]

    \addplot [mark=o, teal!100, line width=0.4mm ] table [x expr=\coordindex, y={encodec_24k_24bps}]{\datatable};


    \addplot [mark=o, teal!60, line width=0.2mm ] table [x expr=\coordindex, y={encodec_24k_3bps}]{\datatable};

    \addplot [mark=square, red!80, line width=0.4mm ] table [x expr=\coordindex, y={funcodec_en_libritts_16k_gr1nq32ds320}]{\datatable};

    \addplot [mark=square, purple!100, line width=0.4mm ] table [x expr=\coordindex, y={speech_tokenizer_16k}]{\datatable};
    

    \addplot [mark=o, blue!80, line width=0.4mm ] table [x expr=\coordindex, y={academicodec_hifi_24k_320d}]{\datatable};


    \addplot [mark=o, orange!100, line width=0.4mm ] table [x expr=\coordindex, y={dac_24khz}]{\datatable};

    \addplot [mark=o, orange!60, line width=0.2mm ] table [x expr=\coordindex, y={dac_24khz_4cb}]{\datatable};

    \addplot [black!80, line width=1.0mm, densely dashed ] table [x expr=\coordindex, y={oracle}]{\datatable};


    
    
    
    
    
    
\end{axis}
\end{tikzpicture}}
    \end{subfigure}

    \caption{Evaluation results of neural audio codecs under different noise conditions. Each row corresponds to a specific noise type (from top to bottom): ambient noise, white noise, and reverberation .}
    \label{fig:noise}
    \vspace{-0.8em}
\end{figure*}
\section{Noise Robustness Evaluation}
Neural audio codecs have proven useful not only for compression but also as tokenizers for language modeling on speech signals. 
However, their deployment in real-world environments—where background noise is common—raises concerns about their robustness.
Understanding how these models perform under noisy conditions is essential for enhancing their reliability in transmission and downstream applications.
In this section, we systematically evaluate their robustness against different types of noise by mixing with clean speech at various noise levels:
\begin{itemize}
    \item \textbf{Ambient Noise}: Simulates real-world environments using everyday background sounds from the WHAM dataset~\cite{wichern2019wham}.
    \item \textbf{White Noise}: Covers the full frequency spectrum, approximating sensor or transmission-induced noise. We generate white noise by sampling from a Gaussian distribution.
    \item \textbf{Reverberation}: Introduces reflections and echoes to simulate reverberant environments. Room impulse responses from the DNS Challenge~\cite{DNS} are used here.
\end{itemize}

Figure~\ref{fig:noise} shows the evaluation results of several neural speech codecs under three noise conditions (from top to bottom): ambient noise, white noise, and reverberation.
Each codec is operated at its highest available bitrate unless otherwise specified.
As an upper-bound reference, we include an \textit{Oracle} topline, obtained by evaluating the noisy (but unencoded) signal directly.
Ideally, a noise-robust codec should exhibit minimal degradation to the topline across conditions and metrics.

Although the perceptual quality (PESQ) of all codecs tends to converge to a similar value in severely noisy conditions, notable differences in their robustness in preserving phonetic content (ASR-WER), speaker characteristics (ASV-EER), and emotion-related acoustic cues (SER-ACC).
When operating at the highest bitrate, DAC consistently demonstrates the strongest noise robustness, exhibiting the least performance degradation across different noise conditions and evaluation metrics, while Encodec and FreqCodec also perform competitively. In contrast, SpeechTokenizer and HiFi-Codec appear significantly more vulnerable to noise.
However, even DAC, the best-performing model, introduces additional distortions under challenging scenarios, notably degrading phonetic content in heavy ambient noise and reducing speaker and emotional fidelity in reverberant environments.
This underscores that noise robustness remains an open challenge for neural speech codecs.
Interestingly, better noise robustness at high bitrates does not necessarily translate to better noise robustness at lower bitrates.
For instance, at 3 kbps, EnCodec outperforms DAC in preserving downstream-related information across most noise conditions.
This finding suggests that noise robustness is not an intrinsic property of high-performing codecs but instead requires specific design considerations.
Additionally, a comparison between HiFi-Codec and EnCodec at similar bitrates reveals that perceptual quality on clean speech does not always correlate with noise robustness, suggesting that training data diversity and quantization strategies might play a crucial role for this characteristic.
Furthermore, frequency-domain quantization, as adapted in FreqCodec, shows promise in preserving emotion-related information, occasionally surpassing higher-bitrate models like EnCodec in this regard.

Overall, our analysis reveals non-trivial discrepancies in noise robustness across neural speech codecs, emphasizing the importance of model design and the need for more rigorously reported evaluations in future research. 
A clear trend emerges: noise robustness is highly sensitive to operating bitrate, which is expected as higher bitrates offer better modeling capacity.
However, the poor noise robustness observed among low-bitrate codecs suggests a key research direction—developing codecs explicitly optimized for noisy conditions, as simply increasing bitrate is not always a feasible solution.

\section{Linearity Analysis}
An ideal speech codec for real-world applications should behave approximately linear to ensure signal integrity when speech overlaps with other sources or when amplitude varies.
Although the underlying neural networks necessarily employ non-linear components, we treat each codec as a black-box system and empirically ask whether its end-to-end input–output mapping satisfies the classical linearity criteria—a prerequisite for lossless compression.
This practical notion of linearity is critical for downstream tasks such as speech separation and enhancement, where non-linear artifacts introduced by the codec can propagate and degrade performance.
Here, we analyze the degree of linearity in neural speech codecs to gain deeper insights into their behavior and to explore potential explanations for their noise robustness performance.
Let $X \to f(X)$ represent the transformation of an input signal $X$ by a system $f(\cdot)$. Then, linearity is characterized by the following properties:
\begin{itemize}
    \item \textbf{Additivity}: $f(X+Y) = f(X)+f(Y)$
    \item \textbf{Homogeneity}: $f(\alpha X) =\alpha f(X)$
\end{itemize}
Additivity ensures the preservation of the structure of overlapping signals, while homogeneity ensures that scaling an input signal yields a proportionally scaled output without distortion.

\begin{figure}[!t]
    \centering
    \vspace{-1.0em}
    \begin{subfigure}{0.48\linewidth}
        \centering
        \resizebox{\linewidth}{!}{


\begin{tikzpicture}
    \begin{axis}[
        title=Additivity,
        title style={font=\large},
        xlabel={Bitrate in kbps},
        ylabel={Mel Distance},
        legend style={at={(0.98,0.98)},anchor=north east},
        grid=major,
        ymajorgrids=true, 
        xmajorgrids=false,
        scatter/classes={
            Encodec_24khz={teal!80, mark=*},
            DAC_24khz={orange!80, mark=*},
            SpeechTokenizer={purple!80, mark=square*},
            HiFiCodec_24khz={blue!80, mark=*},
            FreqCodec={red!80, mark=square*}
        }
    ]

    \addplot[scatter,only marks, mark size=3.5pt,scatter src=explicit symbolic] coordinates {
        (24.0, 0.3973)[Encodec_24khz]
        (24.0, 0.2670) [DAC_24khz]
        (4.0, 0.8201) [SpeechTokenizer]
        (3.0, 1.0204) [HiFiCodec_24khz]
        (16.0, 0.5893) [FreqCodec]
        (12.0, 0.5233)[Encodec_24khz]
        (6.0, 0.6291)[Encodec_24khz]
        (3.0, 0.7088)[Encodec_24khz]
        (12.0, 0.4605)[DAC_24khz]
        (6.0, 0.6048)[DAC_24khz]
        (3.0,0.7171)[DAC_24khz]
    };

    
    \legend{
    EnCodec,
    DAC,
    SpeechTokenizer,
    HiFiCodec,
    FreqCodec
    }

    \addplot[teal!80, line width=0.3mm] coordinates {
        (3.0, 0.7088)
        (6.0, 0.6291)
        (12.0, 0.5233)
        (24.0, 0.3973)
    };

    \addplot[orange!80, line width=0.3mm] coordinates {
        (3.0,0.7171)
        (6.0, 0.6048)
        (12.0, 0.4605)
        (24.0, 0.2670) 
    };


    \end{axis}
\end{tikzpicture}}
    \end{subfigure}
    \hfill
    \begin{subfigure}{0.48\linewidth}
        \centering
        \resizebox{\linewidth}{!}{\pgfplotstableread[col sep=comma,]{homogeneity_mel.csv}\datatable
\begin{tikzpicture}
\begin{axis}[
    title=Homogeneity,
    title style={font=\large},
    xtick=data,
    xticklabels from table={\datatable}{dB},
    x tick label style={font=\normalsize},
    legend style={at={(0.98,0.98)},anchor=north east},
    ylabel={Mel Distance},
    xlabel={Gain (dB)},
    grid=major,
    ymajorgrids=true, 
    xmajorgrids=false,
    ymax=2
    ]


    \addplot [mark=o, teal!100, line width=0.25mm ] table [x expr=\coordindex, y={encodec_24k_24bps}]{\datatable};
    \addlegendentry{EnCodec}

    \addplot [mark=o, orange!80, line width=0.25mm ] table [x expr=\coordindex, y={dac_24khz}]{\datatable};
    \addlegendentry{DAC}

    \addplot [mark=square, purple!100, line width=0.25mm ] table [x expr=\coordindex, y={speech_tokenizer_16k}]{\datatable};

    \addlegendentry{SpeechTokenizer}

    \addplot [mark=o, blue!80, line width=0.25mm ] table [x expr=\coordindex, y={academicodec_hifi_24k_320d}]{\datatable};
    \addlegendentry{HiFi-Codec}


    \addplot [mark=square, red!80, line width=0.25mm ] table [x expr=\coordindex, y={funcodec_en_libritts_16k_gr1nq32ds320}]{\datatable};
    \addlegendentry{FreqCodec}



    
    
    
    
    
    
    
\end{axis}
\end{tikzpicture}}
    \end{subfigure}


    \caption{Linearity analysis of neural speech codecs.}
    \label{fig:linearity}
    \vspace{-0.8em}
\end{figure}
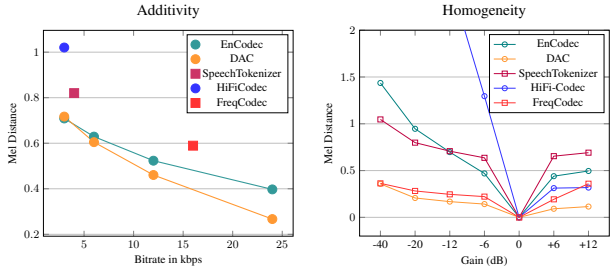
Figure~\ref{fig:linearity} presents our linearity analysis, evaluating additivity across bitrates and homogeneity across gain levels. The additivity results indicate that higher bitrates generally improve additivity, suggesting that existing neural speech codecs do not explicitly optimize for additivity in their training objectives. Meanwhile, the homogeneity analysis reveals distinct behaviors: DAC and FreqCodec remain relatively stable across gain variations, whereas other models exhibit uneven amplification at extreme gains, potentially distorting spectral characteristics.
These trends, in fact, well align with our noise robustness observations: codecs with better additivity (e.g. DAC) introduce fewer non-linear artifacts, leading to stronger noise robustness, while non-homogeneous codecs (e.g. HiFi-Codec) distort spectral dynamics, degrading intelligibility and emotion recognition accuracy.
Overall, this analysis highlights the importance of empirically measured linearity as both an evaluation metric and a potential training objective for future neural speech codecs. 
We encourage research that explicitly regularizes these properties to improve robustness in noisy environments and enhance applicability to tasks involving overlapped signals (e.g., speech separation, speech enhancement).

\section{Frequency Response Characterization}
Speech datasets vary in bandwidth due to differences in recording conditions and applications.
While compression inevitably introduces distortions, a robust neural codec should maintain spectral integrity. 
Evaluating frequency response provides insight into codec behavior and informs potential avenues for improvement. 
In our analysis, we used sine sweeps with a peak amplitude of 1 and validated the results across multiple amplitude levels, consistently observing similar trends.

\begin{figure}
    \centering
\includegraphics[trim={2cm 0.7cm 2cm 4.0cm},width=0.8\linewidth]{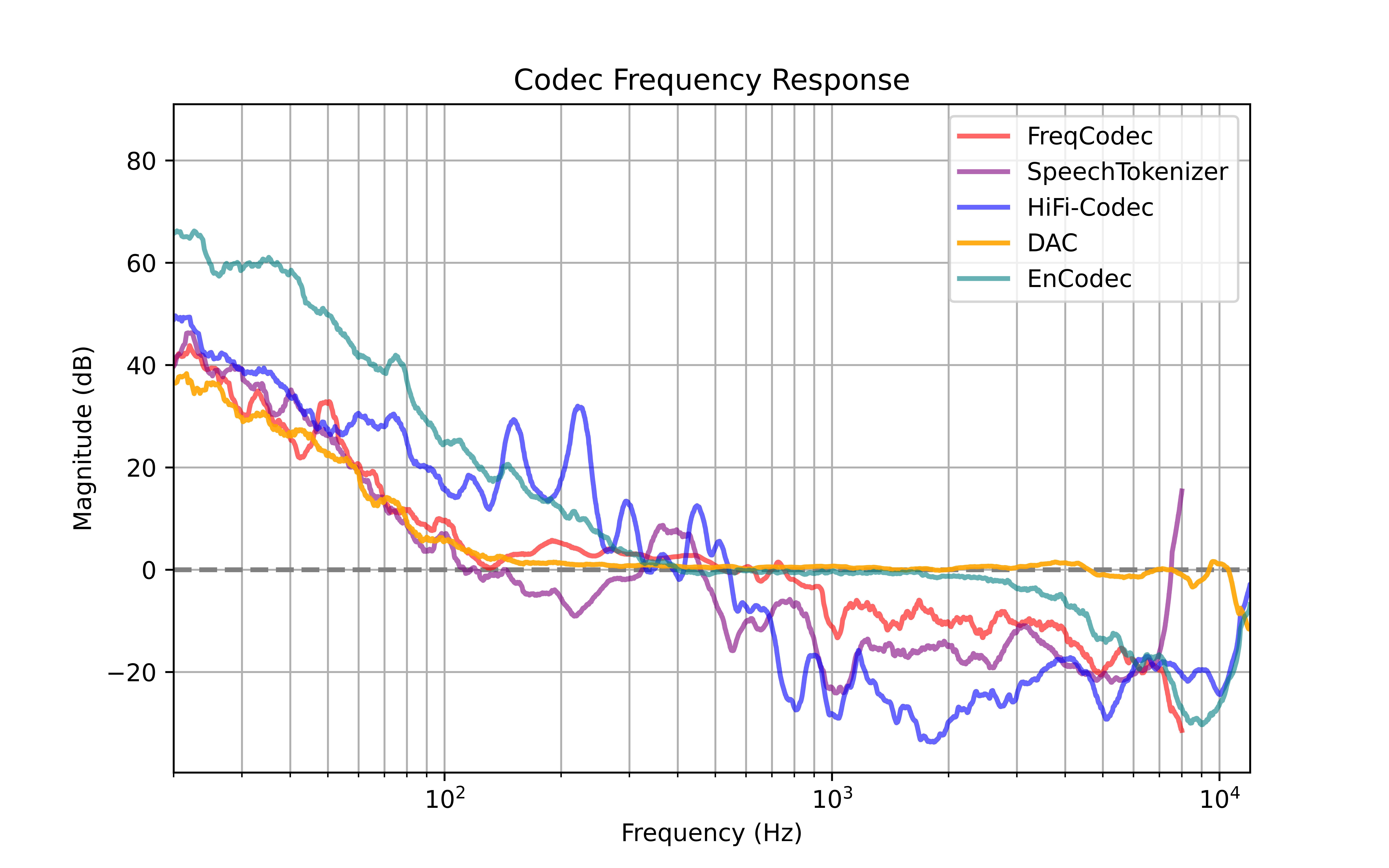}
\vspace{-0.2em}
\caption{Frequency response of neural speech codecs.}
\label{fig:frequency_response}
\vspace{-0.8em}
\end{figure}

Figure~\ref{fig:frequency_response} presents the frequency response of each neural speech codec using sine sweep.
All codecs exhibit low-frequency boosting ($<$100Hz), likely due to the adoption of time-domain loss functions (e.g., L1 loss on waveforms). This effect is most pronounced in EnCodec, which may enhance speech presence but also introduce unintended spectral coloration, potentially causing a ``muddy'' effect.
In the mid-frequency range (100Hz-2kHz), DAC, EnCodec, and FreqCodec maintain a stable response, suggesting better preservation of vocal harmonics and formants, which are essential for phonetic clarity and speaker identity retention. In contrast, HiFi-Codec and SpeechTokenizer exhibit significant fluctuations, which may lead to phoneme distortions, negatively impacting intelligibility (ASR-WER).
At higher frequencies ($>$2kHz), all codecs exhibit attenuation, though DAC and EnCodec demonstrate smoother roll-offs, suggesting better retention of fine-grained details, which may contribute to their higher perceptual scores. Conversely, HiFi-Codec and SpeechTokenizer show more abrupt attenuation, potentially leading to the loss of sibilant sounds and reduced speech clarity.
Overall, this analysis highlights frequency response as a crucial factor in codec performance, emphasizing the need to design models that balance spectral integrity with efficient compression. Future work may explore perceptually weighted loss functions and adaptive spectral processing to further improve speech fidelity.

\section{Conclusions}
In this work, we present a systematic evaluation of neural speech codecs under noisy conditions.
Our findings reveal significant variations in their noise robustness, influenced by factors such as training data diversity, operating bitrate, and quantization strategies.
Additionally, our linearity analysis highlights the presence of non-linear distortions, which may contribute to noise robustness and potential degradation on speech tasks involving overlapped signals.
Lastly, our frequency response characterization further uncovers spectral imbalances, particularly low-frequency emphasis and high-frequency attenuation, which may degrade perceptual quality and downstream task performance.
Our analysis highlights noise robustness as a key challenge in neural speech coding and underscores the need for noise-aware training toward real-world applications and more rigorous evaluation practices in future research.

\bibliographystyle{IEEEtran}
\bibliography{mybib}

\end{document}